\documentclass[11pt,letterpaper]{article}
\usepackage[margin=1in]{geometry}
\usepackage{amsmath,amssymb,amsthm,mathtools}
\usepackage[T1]{fontenc}
\usepackage{lmodern}
\usepackage{enumitem}
\usepackage[expansion=false]{microtype}
\usepackage[colorlinks=true,linkcolor=blue,citecolor=blue,urlcolor=blue]{hyperref}
\newtheorem{theorem}{Theorem}[section]
\newtheorem{lemma}[theorem]{Lemma}
\newtheorem{proposition}[theorem]{Proposition}
\newtheorem{corollary}[theorem]{Corollary}
\theoremstyle{definition}

\theoremstyle{remark}
\newtheorem{remark}[theorem]{Remark}
\newcommand{\F}{\mathbb F}
\newcommand{\Q}{\mathbb Q}
\newcommand{\eps}{\varepsilon}
\newcommand{\rev}{\operatorname{rev}}
\newcommand{\poly}{\operatorname{poly}}
\newcommand{\VPnc}{\mathsf{VP}_{\mathrm{nc}}}
\newcommand{\VNPnc}{\mathsf{VNP}_{\mathrm{nc}}}
\newcommand{\FP}{\mathsf{FP}}
\newcommand{\NP}{\mathsf{NP}}
\newcommand{\SigTwo}{\Sigma_2^{\mathsf P}}

\title{Factor Closure in Noncommutative Algebraic Complexity}
\author{V.\ Arvind\thanks{Institute of Mathematical Sciences, Chennai, India. \texttt{arvind@imsc.res.in}}
\and Pushkar S.\ Joglekar\thanks{Vishwakarma Institute of Technology, Pune, India. \texttt{joglekar.pushkar@gmail.com}}}
\date{November 1, 2026}

\begin{document}
\maketitle

\begin{abstract}
We prove that $\VPnc$ is closed under taking factors over every field. If a noncommutative circuit of size $s$ computes a nonconstant polynomial of degree at most $D$, then each left or right factor has a circuit of size $O(sD^3)$. A central factor $g$ in a decomposition $f=agb$, with nonconstant outer factors $a,b$, has a circuit of size $O(sD^6)$.

The proof uses a small weighted nondeterministic finite automaton (NFA) with replacement labels. Given $f=gh$, its labels depend on a word of maximum degree in $h$ and the coefficients of its suffixes. A pairing of accepting paths proves that the induced linear map $T$ satisfies $T(gh)=g$. Matrix substitution implements $T$ in a circuit. Applying the same substitution to a Boolean-sum verifier proves $\VNPnc$ closure.

The construction also proves formula closure. The automaton has one copying state and, after leaving it, uses only scalar replacements. This structure gives formulas of size $O(s^2)$ for one-sided factors and $O(s^4)$ for central factors of a size-$s$ formula. The same map proves that polynomial-size noncommutative algebraic branching programs are closed under taking factors. Polynomial-size ABP factorizations over finite fields are already known from~\cite{AJ25}.

These are arithmetic size bounds, with arbitrary field constants at unit cost, including over $\Q$. Over finite fields, combining the formula construction with~\cite{AJ22} gives randomized polynomial-time complete factorization of formula inputs, with factors output as formulas. For polynomial-degree circuit inputs, we obtain polynomial-length succinct ABP descriptions with polynomial-time edge queries. We also give two algorithms that output complete ordered irreducible factorizations as polynomial-size circuits: a deterministic $\FP^{\SigTwo}$ algorithm and a zero-error randomized algorithm with an NP oracle and expected polynomial running time.
\end{abstract}
\newpage

\section{Introduction}

Let $f\in\F\langle X\rangle$ be a nonzero polynomial of degree at most $D$, computed by a circuit of size $s$. We study the circuit size of a specified factor $g$ in an identity $f=agb$. The objective is a bound polynomial in $s$ and $D$, with no homogeneity assumption and over an arbitrary field. We also study the corresponding questions for formulas, algebraic branching programs, and polynomials presented as Boolean sums of small circuits. The bounds must hold for every specified factorization, since noncommutative factorizations need not be unique.

Over finite fields, polynomial-time factorization is known for ABP inputs, with the factors output as ABPs~\cite{AJ25}. That work explicitly leaves the circuit-input problem open: its concluding remarks identify Higman linearization as an obstacle to extending the algorithm and report no nontrivial complexity upper bound for general circuit-input factorization. A basic structural question precedes the algorithmic one: do factors of a small circuit even have small circuits? We answer this question for polynomial-degree inputs. Our formula theorem also preserves the more restrictive input model, giving small formulas for factors; combined with~\cite{AJ22}, it yields finite-field factorization with formula outputs. The oracle algorithms later in the paper use the circuit-size bound to search for factors, while ordinary polynomial-time factorization of polynomial-degree circuit inputs remains open.

Closure also connects factorization to lower bounds. If a polynomial family requires large circuits, and small circuits are closed under taking factors, then every nonzero multiple of that family of polynomially bounded degree requires large circuits as well. For a class defined by a sum over Boolean assignments, closure asks a related question: does a factor still admit a short circuit whose Boolean sum gives the polynomial? These questions concern the existence of small representations. Finding such representations from the product alone is a further algorithmic problem.

In the commutative setting, small-circuit bounds for factors were established in~\cite{Kal89}. Closure of commutative $\mathsf{VNP}$ in characteristic zero and sufficiently large characteristic was proved in~\cite{CKS19}. More recently, commutative formula closure in characteristic zero was proved in~\cite{BKRSS25}. We study the corresponding questions in the free associative algebra $\F\langle X\rangle$. Here monomials are words, and multiplication concatenates them in order. Factorizations need not be unique. We consequently formulate our results for every specified factor: whenever $f=agb$ with all three polynomials nonzero, we bound the complexity of $g$.

\subsection{Main results}

Our circuit model permits arbitrary constants from $\F$, with each constant counted as one input gate. A left or right factor is called one-sided. We call $g$ a central factor in a decomposition $f=agb$ when both outer factors $a,b$ are nonconstant. This terminology describes its position in the specified decomposition; the same polynomial can also occur as a one-sided factor in another decomposition. All circuit-size assertions below use this arithmetic model.

\begin{theorem}[Circuits for factors]\label{thm:size}
Let a noncommutative circuit of size $s$ compute a nonzero polynomial $f$ of degree at most $D$ over a field $\F$, where $D\ge1$.
\begin{enumerate}[label=(\roman*),leftmargin=*]
\item If $f=gh$ with $g,h\ne0$, then each of $g,h$ has a circuit of size $O(sD^3)$.
\item If $f=agb$ with $a,g,b\ne0$, then $g$ has a circuit of size $O(sD^6)$. This covers central factors as well as one-sided factors.
\end{enumerate}
The circuits can be constructed deterministically with $\poly(s,D)$ field operations when the following cofactor data are supplied. To recover $g$ from $f=gh$, supply a maximum-degree word of $h$, its nonzero coefficient, and the coefficients of all its proper suffixes, including the empty suffix. To recover $h$, supply the analogous data for $\rev(g)$. To recover $g$ from $f=agb$, supply these data for $b$ and for $\rev(a)$. Thus the central-factor construction uses two words of length at most $D$ and $O(D)$ field elements. A scalar cofactor requires only its nonzero value. The construction outputs a circuit for the specified factor and preserves any syntactic degree bound on the input circuit.
\end{theorem}

\begin{theorem}[Formula bounds and closure]\label{thm:closure}
Over any field, polynomial-size noncommutative formulas, $\VPnc$, and $\VNPnc$ are closed under taking factors. In each case this means closure for every specified $g_n$ in $f_n=a_ng_nb_n\ne0$.

More precisely, a one-sided factor of a polynomial computed by a size-$s$ formula has a formula of size $O(s^2)$. A central factor $g$ in $f=agb\ne0$ has a formula of size $O(s^4)$.
\end{theorem}

For a constant $f$, all its factors are nonzero constants and have constant-size circuits. The degree bound in Theorem~\ref{thm:size} concerns the output polynomial; intermediate gates may have larger degree. The construction assumes that the stated cofactor data are supplied and does not find them from the input circuit. The theorem applies to nonhomogeneous polynomials and to every characteristic. It also applies over $\Q$: a bound on the number of additional constants is enough for arithmetic circuit size. Bounds on their binary lengths are a separate question, discussed in Section~\ref{sec:fields}.

We also obtain algorithmic consequences over explicitly represented finite fields. Every factor of a polynomial-degree circuit has a polynomial-length succinct ABP description: it specifies the layers and a polynomial-time algorithm that returns the affine edge label at any requested pair of vertex indices. The ABP itself may have exponential width. Given the input circuit and a degree bound in unary, our factorization algorithms output a scalar and an ordered list of circuits for nonconstant irreducible factors. They run deterministically in $\FP^{\SigTwo}$, or with zero error and expected polynomial time using an NP oracle. The randomized algorithm first samples a list of identity tests and uses the oracle to certify that the list works for all bounded candidate circuits. For formula inputs, combining our formula construction with~\cite{AJ22} instead outputs the irreducible factors as formulas in randomized polynomial time.

\subsection{The idea of the proof}

Suppose $f=gh$. Choose a word $v=v_1\cdots v_b$ of maximum length in $h$, with coefficient $c\ne0$. Our weighted NFA first copies a chosen prefix of its input. It then follows the letters of $v$, replacing them by $1$, with additional transitions that return to the start of this part and replace a letter by a negative suffix coefficient divided by $c$.

Fix a word $r$ in place of $g$. Only suffixes of $v$ in $h$ can contribute to $T(rh)$. Every accepting path on $rv$ that returns to state $0$ has a partner on a shorter input: delete its last return block. The two paths have opposite contributions. The sole unpaired path copies $r$ and erases $v$, proving $T(rh)=r$. Linearity gives $T(gh)=g$.

Replacing each variable by a transition matrix applies this map to a circuit. The same substitution in a VNP verifier works because $T$ commutes with its finite Boolean sum. For formulas, we use one further fact: the matrix block that describes erasing contains only field constants. At a product gate, these constants can be evaluated first, so there is no need to duplicate a polynomial formula for every matrix entry. This gives the quadratic formula bound.

\subsection{Relation to earlier work}

Variable-disjoint, homogeneous, and multilinear noncommutative factorization are studied in~\cite{AJR18}. That work relates variable-disjoint factorization to polynomial identity testing and reduces homogeneous factorization to that case by labelling variables with their positions. The general-circuit algorithms use identity testing and output circuits; the ABP versions admit deterministic polynomial-time algorithms with ABP outputs.

The closest connection with our proof is word differentiation. If $h$ is homogeneous and $[v]h=c\ne0$ for a degree-$b$ word $v$, retaining the words of $gh$ that end in $v$ and removing that suffix gives $cg$. Our automaton does exactly this when all reset labels vanish. For a nonhomogeneous $h$, simply removing $v$ can also retain contributions from shorter words of $h$ completed by letters from $g$. The suffix coefficients supply reset labels that cancel these additional contributions. Thus the additional step is a linear extraction identity for arbitrary nonhomogeneous cofactors, proved in Section~\ref{sec:nfa}; the implementation by automata and matrix substitution builds on~\cite{AMS10,AJR18,AJMR19}.

There is also an algorithmic distinction. The algorithms in~\cite{AJR18} find factors in their stated settings. Our circuit-closure proof constructs a specified factor once a longest cofactor word and its suffix coefficients are supplied; it does not find those data from a general circuit in polynomial time. Homogeneity removes the shorter-suffix terms, which is why the homogeneous extraction needs less information.

Computing the homogeneous components of $f$ alone does not resolve this issue. Writing $g^{(i)}$ and $h^{(j)}$ for the homogeneous components of the factors, one has $f^{(k)}=\sum_{i+j=k}g^{(i)}h^{(j)}$. A component of the product is generally a sum of products, so the homogeneous factorization result does not directly recover the nonhomogeneous factors. Our map acts on the whole polynomial $f$ and needs only one longest cofactor word and its suffix coefficients.

The algorithm of~\cite{AJ22} takes finite-field formula inputs and outputs factors as ABPs. The journal version~\cite{AJ25} extends this to ABP inputs, so polynomial-size ABP factorizations over finite fields are already known. These algorithms use Higman linearization and invariant-subspace methods. We include a direct ABP closure proof to show how the same extraction map acts on this model. The formula result gives a stronger output guarantee for formula inputs: using the cofactor ABPs supplied by~\cite{AJ22}, our construction produces formulas for the factors. Section~\ref{sec:formula-algorithm} explains this combination. The new formula construction uses the existing algorithm to find the factorization.

The distinction between these representations matters for the size bounds. The ABP algorithm measures its cost and output size in the input ABP size; it does not supply a polynomial-size ABP for an arbitrary small circuit. Conversely, its ABP outputs do not by themselves establish formula closure. For circuits, our extraction map is applied directly to the input circuit. For formulas, we additionally use the automaton's scalar erasing block to avoid duplicating polynomial subformulas for all matrix entries. These are the two size-preservation arguments behind our main results.

Factorization using minimal linear representations is studied in~\cite{Sch19}. Our bounds are measured in the size of the original circuit, which can be much smaller than an explicit linear representation.

Our formula result also complements the characteristic-zero commutative formula-closure theorem of~\cite{BKRSS25}. The proof here uses the order of letters in a noncommutative word: the automaton copies a prefix and cancels the unwanted suffix contributions. Its scalar erasing block then gives a small formula for the extracted factor. Both steps work in every characteristic.

\section{Polynomials, circuits, and fields}\label{sec:prelim}

Let $X=\{x_1,\ldots,x_m\}$, with $m\ge1$, and let $X^*$ be the set of finite words in $X$. The empty word $\eps$ represents the multiplicative identity. For a polynomial $P$ and a word $w$, write $[w]P$ for the coefficient of $w$. Multiplication satisfies
\begin{equation}\label{eq:product-coeff}
 [w](PQ)=\sum_{uv=w}[u]P\,[v]Q.
\end{equation}
The degree of a nonzero polynomial is the maximum length of a word with nonzero coefficient.

\begin{lemma}\label{lem:degree}
If $P,Q\ne0$, then $PQ\ne0$ and $\deg(PQ)=\deg P+\deg Q$.
\end{lemma}
\begin{proof}
Put $a=\deg P$ and $b=\deg Q$. Choose words $u,v$ of lengths $a,b$ with nonzero coefficients in $P,Q$, respectively. A contribution to $[uv](PQ)$ must use a word of length at most $a$ from $P$ and one of length at most $b$ from $Q$. Since $|uv|=a+b$, both lengths must be maximal. The cut is therefore fixed after the first $a$ letters, and $[uv](PQ)=[u]P[v]Q\ne0$. No product word is longer than $a+b$.
\end{proof}

A factor of a nonzero polynomial $f$ means a polynomial $g$ such that $f=agb$ for some $a,b\ne0$. A left factor has a presentation $f=gh$, and a right factor has a presentation $f=hg$. In a presentation $f=agb$ with both $a,b$ nonconstant, we call $g$ a central factor. These descriptions need not be exclusive: they refer to positions in a chosen decomposition. Nonzero scalars are units.

A nonconstant polynomial is irreducible if it is not a product of two nonconstant polynomials. By Lemma~\ref{lem:degree}, repeated splitting gives at most $\deg f$ nonconstant irreducible factors. For two or more variables, ordered factorizations into irreducibles need not be unique up to scalar rescaling~\cite{AJR18}. Our closure results apply to every specified $g$ in every identity $f=agb\ne0$; they do not select a preferred factorization. The algorithms later in the paper output one complete factorization $f=\lambda f_1\cdots f_k$, consisting of a nonzero field element $\lambda$ and an ordered list of nonconstant irreducible factors in the stated representation. The list is empty for a scalar input.

\paragraph{Circuit conventions.}
Circuits have variable or field-constant input gates, binary addition gates, and ordered binary multiplication gates. We count gates. A formula is a circuit whose underlying computation is a tree; repeated uses of a subexpression are counted separately. There are no division gates. Field inverses used in constructing a circuit may be supplied as constant inputs. Syntactic degree is zero at a constant, one at a variable, the maximum at an addition gate, and the sum at a multiplication gate.

For $D\ge1$, an actual degree bound $D$ can be converted into a syntactic bound with polynomial overhead. At every gate retain the homogeneous components of degrees $0,\ldots,D$. Addition is componentwise, and multiplication uses
\[
 (PQ)^{(d)}=\sum_{i=0}^d P^{(i)}Q^{(d-i)}.
\]
Components of higher degree cannot contribute to a lower degree and can be discarded. Summing the output components gives a circuit of size $O(sD^2)$ and syntactic degree at most $D$ for any size-$s$ circuit whose output degree is at most $D$. If $D=0$, evaluating the input circuit at $X=0$ gives its constant output in $O(s)$ field operations.

\paragraph{Reversal.}
Reverse the order of letters in a word and extend this operation linearly, writing it as $\rev$. Then $\rev(PQ)=\rev(Q)\rev(P)$. Reversing the order of the two inputs at every multiplication gate gives a same-size circuit for the reversed polynomial.

\subsection{Arithmetic size and binary length}\label{sec:fields}

In the arithmetic model, any element of $\F$ may appear as a constant input. Its cost is one gate, independent of how it is represented. All our arithmetic closure statements use this model and are nonuniform: a different circuit may be chosen for each input length. These conventions are the usual ones for the classes~\cite{AJR16}.

In particular, a rational number is allowed at one input gate even if its numerator or denominator has many bits. The NFA below is specified by the additional constants
\[
 c^{-1}\quad\hbox{and}\quad \frac{[t_\ell]h}{c}\quad(1\le\ell\le\deg h),
\]
where $c$ is a nonzero top-degree coefficient and the $t_\ell$ are its word's proper suffixes. There are only linearly many such constants. The formula construction also evaluates scalar matrix expressions in these and the input constants, and puts the resulting field elements at its leaves. The number of leaves remains polynomial. Binary lengths play no role in these arithmetic size bounds. Thus a coefficient-length bound for factors over $\Q$ is not an assumption of the closure theorems.

Binary descriptions require a different qualification. The present proof does not bound the binary lengths of these rational constants in terms of the original circuit alone. Moreover, an exact factor can be rescaled by any nonzero rational $\lambda$, with the inverse scalar absorbed into an adjacent factor. A claim about binary descriptions of all exact factors must account for this choice. We do not claim closure with polynomial-bit constants, or with constants restricted to a fixed finite set.

For circuit construction over $\Q$, a conditional bit bound is available. Suppose the valid cofactor words and coefficients required by Theorem~\ref{thm:size} are supplied explicitly, with total binary length $H$. The construction then outputs the factor circuit in time polynomial in the encoded input size, $D$, and $H$. The required inverses and coefficient ratios have polynomial bit length, and the remaining operations create circuit gates. This statement neither bounds $H$ in terms of the original circuit nor provides an algorithm for finding the cofactor data. It also does not assert a bit bound for the constants evaluated in the formula construction.

All subsequent algorithmic bit bounds use an explicitly represented finite field $\F_q=\F_p[t]/(\varphi(t))$, with the prime $p$ and an irreducible polynomial $\varphi$ supplied in a valid encoding. A field element uses $O(\log q)$ bits, and field operations take polynomial time in $\log q$. Degree bounds supplied to the circuit-factorization algorithms are encoded in unary. Output factors are presented by circuits or formulas over the same field, as specified in each result.

\section{An NFA that removes a right cofactor}\label{sec:nfa}

Fix $h\ne0$ with $b=\deg h\ge1$. Choose $v=v_1\cdots v_b$ with $c=[v]h\ne0$. For $1\le\ell\le b$, put
\begin{equation}\label{eq:data}
 p_\ell=v_1\cdots v_\ell,\qquad
 t_\ell=v_{\ell+1}\cdots v_b,\qquad
 d_\ell=[t_\ell]h,\qquad
 \alpha_\ell=d_\ell/c.
\end{equation}
Thus $v=p_\ell t_\ell$ and $t_b=\eps$. These data will determine a linear map $T$. They are chosen from $h$; no information about the polynomial multiplied on its left is needed.

\subsection{States and replacement labels}

The machine has states $*,0,1,\ldots,b$. Its initial state is $*$, and its only accepting state is $b$. Each transition specifies the input letter to read and the expression by which to replace it. The following are all transitions; the last two rows apply for $0\le i<b$.
\begin{center}
\renewcommand{\arraystretch}{1.35}
\begin{tabular}{c|c|c}
Transition & Input consumed & Replacement\\ \hline
$*\longrightarrow *$ & any $x\in X$ & $x$\\
$*\longrightarrow 0$ & $\eps$ & $1$\\
$i\longrightarrow i+1$ & $v_{i+1}$ & $1$\\
$i\longrightarrow 0$ & $v_{i+1}$ & $-\alpha_{i+1}$
\end{tabular}
\end{center}
The transition with input $\eps$ consumes no letter. The last row is a \emph{reset}. State $b$ has no outgoing transitions, and no state $0,\ldots,b$ can return to $*$. An accepting path must consume the whole input and end at $b$.

For a word $w$, multiply the replacement labels along each accepting path, in their order. Sum these products and multiply by $c^{-1}$; this defines $T(w)$. Extend the definition linearly:
\begin{equation}\label{eq:T-definition}
 T(P)=\sum_{w\in X^*}[w]P\,T(w).
\end{equation}
If $P=\sum_w a_w w$, an accepting run on $w$ with ordered replacement product $Q$ contributes $a_wc^{-1}Q$ to $T(P)$. Summing these products over all runs and all input words gives~\eqref{eq:T-definition}. At $*$ the machine copies letters. After leaving $*$ it replaces each letter by a field element, so $Q$ is a scalar multiple of the copied prefix.

All sums are finite. The empty-input transition can be used at most once, and every other transition consumes a letter. Only finitely many input words have nonzero coefficient in $P$. Hence $T$ is a well-defined linear map and never increases degree. We keep transitions with replacement $0$ as formal paths contributing zero.

\subsection{The cancellation proof}

A reset block of length $\ell$ starts at $0$, advances $\ell-1$ times, and then resets to $0$. It reads $p_\ell$ and replaces this whole block by $-\alpha_\ell$. The case $\ell=1$ is a loop at $0$. Every accepting path ends with a full advancing traversal of $v$: this traversal begins after its last reset, or just after leaving $*$ if there is no reset. In particular, every accepted word ends in $v$.

\begin{theorem}[Removal of a right cofactor]\label{thm:remove}
For every word $r$, $T(rh)=r$. Consequently, for every polynomial $g$,
\[
 T(gh)=g.
\]
Equivalently, $[u]T(gh)=[u]g$ for every word $u\in X^*$, including $u=\eps$, whose coefficient is the constant term.
\end{theorem}
\begin{proof}
Fix $r$. Every word $s$ in $h$ has length at most $b$. If $rs$ has an accepting path, it ends in $v$, so $s$ must be the suffix of $v$ of length $|s|$. Therefore
\begin{equation}\label{eq:rh-suffix}
 T(rh)=cT(rv)+\sum_{\ell=1}^b d_\ell T(rt_\ell).
\end{equation}
Terms with $d_\ell=0$ are included formally.

The word $r$ may end in several of the words $p_1,\ldots,p_b$. We do not choose one of these suffixes in advance. Instead, fix one accepting path on $rv$ that has a reset. This path determines a unique index $\ell$: its last reset goes from state $\ell-1$ to state $0$. The block ending at that reset begins immediately after the preceding reset, or immediately after leaving $*$ if there is no preceding reset. It advances $\ell-1$ times and then resets, so it reads exactly $p_\ell$. The final advancing traversal of $v$ follows this block. Let $w$ be the input read before that reset block, and let $B$ be the product of all replacement labels on this earlier part of the path. This part ends at state $0$. The input has the form
\[
 rv=wp_\ell v.
\]
Cancelling the common suffix $v$ gives $r=wp_\ell$, and therefore
\begin{equation}\label{eq:paired-input}
 rt_\ell=wp_\ell t_\ell=wv.
\end{equation}
Keep the entire earlier path on $w$, including its choice of when to stop copying and all its earlier resets, and delete the last reset block. Follow it directly by the advancing traversal of $v$. This is an accepting path on $rt_\ell$.

Including the coefficients in~\eqref{eq:rh-suffix} and the final multiplier $c^{-1}$, the two paths contribute
\[
 c\,B(-\alpha_\ell)c^{-1}=-\alpha_\ell B,
 \qquad
 d_\ell Bc^{-1}=+\alpha_\ell B.
\]
They cancel. The polynomial $B$ is a scalar multiple of a copied prefix, so cancellation occurs at that same output monomial. The paired input words can be different: their lengths differ by $\ell$.

To check that every suffix-term path is paired exactly once, start with an accepting path on $rt_\ell$. Write its input as $wv$, where the final $v$ is its last advancing traversal. The path before that traversal ends at $0$. Since
\[
 rt_\ell=wv=wp_\ell t_\ell,
\]
right cancellation gives $r=wp_\ell$. Insert a reset block reading $p_\ell$ immediately before the final traversal of $v$. The resulting path reads $rv$, and the inserted block is its last reset. This reverses the deletion uniquely. The argument also applies when $\ell=b$ and $t_\ell=\eps$.

For each fixed $\ell$, deletion and insertion therefore give inverse correspondences between the paths on $rv$ whose last reset has index $\ell$ and all paths in the summand $d_\ell T(rt_\ell)$. The first collections are disjoint for different $\ell$: one path has only one last reset. Several $p_\ell$ can be suffixes of $r$, and several different paths can have the same last reset index, but every individual path still occurs in exactly one pair. The earlier path is preserved, so distinct histories are not identified. In particular, no path is cancelled twice. The copied prefix is also preserved, proving cancellation separately at each output monomial.

Only the accepting path on $rv$ with no reset remains. It must copy exactly $r$ and then replace the final $v$ by $1$. Its contribution is $cc^{-1}r=r$. Zero labels cause no problem, and the cancellation is valid in every characteristic. We have proved, for every word $u$,
\begin{equation}\label{eq:word-coeff}
 [u]T(rh)=
 \begin{cases}1,&u=r,\\0,&u\ne r.\end{cases}
\end{equation}
Finally, write $g=\sum_r a_r r$. Linearity gives
\[
 T(gh)=\sum_r a_rT(rh)=\sum_r a_rr=g,
 \qquad [u]T(gh)=a_u=[u]g.
\]
\end{proof}

The proof expands $gh$ to identify cancelling contributions. The machine can instead act directly on the combined coefficients of $f=gh$: multiplication by a path's replacement product distributes over the finite sum $[w]f=\sum_{rs=w}[r]g[s]h$. No grouping of monomials is needed.

If $h=c\ne0$ is a scalar, use one initial and accepting state, with a loop reading and replacing $x$ by $x$ for every variable, and multiply the total by $c^{-1}$. Then $T(P)=c^{-1}P$.

\begin{remark}[The homogeneous case]\label{rem:homogeneous}
If $h$ is homogeneous of degree $b\ge1$, then all $d_\ell$ in~\eqref{eq:data} are zero. Only paths without resets can contribute nonzero terms, and
\[
 T(P)=c^{-1}\sum_u[uv]P\,u.
\]
This is the familiar operation that removes a fixed word from the right. For a nonhomogeneous $h$, the proper suffixes can have nonzero coefficients. The reset transitions provide exactly the cancellations proved above. This explains the additional step beyond the homogeneous word-removal argument, while keeping the same matrix-substitution method.
\end{remark}

\section{Implementing the NFA in a circuit}\label{sec:matrix}

The use of finite automata to define matrix substitutions for noncommutative circuits appears in~\cite{AMS10}. Substitution automata, whose transitions replace input letters by variables or scalars, are used in~\cite{AJR18,AJMR19}. We apply this established implementation method to the factor-extraction automaton of Theorem~\ref{thm:remove}. Evaluating the input circuit on its transition matrices sums the accepting-path contributions without expanding the circuit into monomials.

For each $x\in X$, let $A_x$ be the $(b+1)\times(b+1)$ scalar transition matrix on the erasing states $0,\ldots,b$. Its $(i,j)$ entry is the sum of the replacement labels on transitions from $i$ to $j$ reading $x$. Thus, for each $0\le i<b$ with $x=v_{i+1}$, add $1$ at $(i,i+1)$ and $-\alpha_{i+1}$ at $(i,0)$. All other entries are zero. Write $e_i$ for the coordinate column vector of state $i$.

Index an additional row and column by $*$ and define
\begin{equation}\label{eq:M}
 M_x=\begin{pmatrix}
 x&e_0^{\mathsf T}A_x\\
 0&A_x
 \end{pmatrix}.
\end{equation}
These matrices have entries in $\F\langle X\rangle$. The top-left entry copies $x$. The row $e_0^{\mathsf T}A_x$ combines the empty-input transition $*\to0$ with the first letter read after that transition. Thus the matrices have no empty-input edges to handle.

\begin{lemma}[Circuit implementation]\label{lem:matrix}
For the nonscalar cofactor map, with $b\ge1$, every polynomial $P$ satisfies
\begin{equation}\label{eq:T-matrix}
 T(P)=c^{-1}\bigl[P((M_x)_{x\in X})\bigr]_{*,b}.
\end{equation}
If $P$ has a circuit of size $s$, then $T(P)$ has a circuit of size $O(sb^3)$. The construction preserves any syntactic degree bound on the given circuit. Scalar cofactor removal uses one scalar multiplication.
\end{lemma}
\begin{proof}
Let $w=x_{i_1}\cdots x_{i_k}$ be nonempty. Multiplying the block matrices gives
\begin{equation}\label{eq:block}
 [M_{x_{i_1}}\cdots M_{x_{i_k}}]_{*,b}
 =\sum_{j=0}^{k-1}x_{i_1}\cdots x_{i_j}\,
 e_0^{\mathsf T}A_{x_{i_{j+1}}}\cdots A_{x_{i_k}}e_b.
\end{equation}
The term indexed by $j$ copies the first $j$ letters. The scalar matrix product sums all erasing paths on the remaining suffix, starting at $0$ and ending at $b$. It therefore accounts for exactly the runs that switch from copying to erasing after $j$ letters. An accepting run cannot switch after the last letter: reaching $b$ from $0$ requires further letters because $b\ge1$. These are all accepting runs on $w$.

For $w=\eps$, matrix substitution gives the identity matrix, whose $(*,b)$ entry is zero. The NFA has no accepting run on $\eps$ either. Multiplying~\eqref{eq:block} by $c^{-1}$ and extending linearly proves~\eqref{eq:T-matrix}.

To obtain a circuit, replace a scalar input $a$ by $aI_{b+2}$ and a variable input $x$ by $M_x$. Replace every internal gate by matrix addition or ordered matrix multiplication. Express each matrix entry using scalar gates. Addition requires $O(b^2)$ gates and multiplication requires $O(b^3)$ gates per original gate. Previously computed entries are shared. Finally take the designated entry and multiply by $c^{-1}$.

At a scalar input every entry has syntactic degree zero; at a variable input it has syntactic degree at most one. Induction over the gates proves that each new entry has syntactic degree at most that of its original gate. No division gate is introduced.
\end{proof}

\begin{proof}[Proof of Theorem~\ref{thm:size}]
For $f=gh$ with $\deg h\ge1$, combine Theorem~\ref{thm:remove} and Lemma~\ref{lem:matrix}. This gives a circuit for $g$ of size $O(s(\deg h)^3)\subseteq O(sD^3)$. If $h$ is a scalar, multiply the input circuit by $h^{-1}$.

For $h$, reverse the circuit for $f$. Since $\rev(f)=\rev(h)\rev(g)$, the same construction gives $\rev(h)$; reverse the resulting circuit to obtain $h$.

For a factorization $f=agb$, first remove the right cofactor $b$ to obtain $ag$. Then use reversal to remove the left cofactor $a$ and obtain $g$. By Lemma~\ref{lem:degree}, both cofactor degrees are at most $D$. There are at most two matrix substitutions, with scalar multiplications for scalar cofactors, so the size is $O(sD^6)$. Each substitution needs one word and linearly many field elements, and preserves syntactic degree.

The structural estimate requires only the actual degree of $f$. Substitution into matrices respects addition and multiplication, so any cancellations already present in the input circuit remain valid after substitution, even if some intermediate gates have very large degree.
\end{proof}

\begin{remark}[A common bound for all factors]\label{rem:bound}
For an input circuit of syntactic degree at most $D$, where $D\ge1$, fix an absolute constant $K\ge1$ large enough that
\[
 B(s,D)=KsD^6
\]
bounds the above construction. Every factor has a circuit of size at most $B(s,D)$ and syntactic degree at most $D$. Crucially, this is a bound in terms of the original circuit; it need not grow each time a factor is split further.
\end{remark}

\section{Formulas and algebraic branching programs}\label{sec:smaller-models}

\subsection{A direct formula construction}\label{sec:formulas}

We apply the factor-extraction NFA of Section~\ref{sec:nfa} to a polynomial $P$ computed by a formula. As in Section~\ref{sec:matrix}, this means substituting its transition matrix $M_x$ from~\eqref{eq:M} for each variable $x$. The substitution produces the matrix $P(M)$, and the required polynomial is its normalized entry
\[
 T(P)=c^{-1}[P(M)]_{*,b}.
\]
We will construct a small formula for this entry directly.

The NFA first copies letters and then replaces letters by scalars. It can change from copying to replacing only once. Correspondingly, the upper-left entry of $P(M)$ is $P$, and the lower-right block contains only field elements. The latter block can therefore be evaluated before constructing the output formula. At a multiplication gate, the change can occur in the left part of the input word or in the right part. These two possibilities give the recursion below.

\begin{lemma}[Applying the map to a formula]\label{lem:formula}
If $P$ has a formula of size $s$, then $T(P)$ has a formula of size $O(s^2)$. For $b\ge1$, given the NFA data, the formula can be constructed using $O(sb^3)$ field operations and $O(s^2)$ operations to write the formula. A scalar cofactor requires only scalar multiplication of the original formula.
\end{lemma}
\begin{proof}
Assume $b\ge1$; a scalar cofactor only requires a scalar multiplication. Let $\Phi$ be the original formula, with root $\rho$. For each node occurrence $U$ of $\Phi$, let $p_U$ denote the polynomial computed by the original subtree rooted at $U$. Thus $p_\rho=P$. Throughout this proof, $V,W$ denote the left and right children of an internal node $U$.

Recall that $M_x$ has blocks $x$, $e_0^{\mathsf T}A_x$, $0$, and $A_x$, where $A_x$ is a $(b+1)\times(b+1)$ matrix over $\F$. Evaluation of the subtree at $U$ gives
\begin{equation}\label{eq:formula-block}
 p_U(M)=\begin{pmatrix}p_U&R_U\\0&C_U\end{pmatrix}.
\end{equation}
Here $R_U$ is a row of $b+1$ noncommutative polynomials, and $C_U$ is a $(b+1)\times(b+1)$ matrix over $\F$. Concretely, to compute $C_U$, take the original subtree at $U$, replace each variable leaf $x$ by $A_x$ and each scalar leaf $a$ by $aI_{b+1}$, and perform its additions and ordered multiplications as matrix operations. No polynomial variables remain in this evaluation. Thus $C_U=A_x$ at a variable leaf, $C_U=aI_{b+1}$ at a scalar leaf, and at internal nodes
\[
 C_U=C_V+C_W\quad\text{or}\quad C_U=C_VC_W,
\]
according to the gate. We compute all these matrices from the leaves to the root.

\emph{Constructing the formula.}
For a column $z\in\F^{b+1}$, let $F(U,z)$ be the formula we construct to compute the single polynomial $R_Uz$. The column consists of field elements, not polynomial variables. At a scalar leaf $U$ labelled $a$, the row $R_U$ is zero, so $F(U,z)$ is the single constant leaf $0$. At a variable leaf labelled $x$, it is the single constant leaf $e_0^{\mathsf T}A_xz$.

At an addition node $U$, construct
\begin{equation}\label{eq:formula-add}
 F(U,z)=F(V,z)+F(W,z).
\end{equation}
At a multiplication node $U$, construct
\begin{equation}\label{eq:formula-product}
 F(U,z)=p_VF(W,z)+F(V,C_Wz).
\end{equation}
In this rule, the factor $p_V$ is represented by an \emph{unchanged copy of the original subtree at $V$}. Only the two occurrences of $F$ are recursive calls. The scalar column $C_Wz$ is computed before making the call on $V$.

To verify the product rule, multiply the two block matrices:
\[
 p_U(M)=p_V(M)p_W(M)
 =\begin{pmatrix}
 p_Vp_W&p_VR_W+R_VC_W\\
 0&C_VC_W
 \end{pmatrix}.
\]
Consequently, $R_Uz=p_V(R_Wz)+R_V(C_Wz)$, exactly as in~\eqref{eq:formula-product}. The first term corresponds to copying the whole left part and changing to scalar replacement in the right part. The second corresponds to changing in the left part; the right part then acts only through the matrix $C_W$ over $\F$. The addition rule follows from block addition. Induction on the original formula proves that $F(U,z)$ computes $R_Uz$. In particular, the desired output is the formula $c^{-1}F(\rho,e_b)$.

\emph{Why the NFA structure matters.}
Two features of~\eqref{eq:formula-block} make this recursion work. The upper-left block is the single original polynomial $p_V$, so its occurrence in~\eqref{eq:formula-product} is supplied by an unchanged subtree. The lower-right block $C_W$ has entries in the field, so $C_Wz$ is again a column of field elements. Thus the recursive call has exactly the same form: its column parameter carries no polynomial formulas, and every leaf contraction is just one precomputed constant. This permits one recursive call on each original child. A general NFA with replacement labels need not have this triangular form or a lower block over the field. If $C_W$ contained polynomial entries, $C_Wz$ could no longer be precomputed as field data, and the leaf contractions would themselves require polynomial formulas. The size recurrence below would then no longer follow. The argument therefore uses the particular structure of our NFA; it does not establish the same bound for an arbitrary NFA.

\emph{Formula size.}
Write $|U|$ for the number of nodes in the original subtree at $U$, including leaves, so $|\rho|=s$. Let $n(U)$ be the number of nodes in the unsimplified formula $F(U,z)$. This count is independent of $z$: each evaluated constant, even zero, is kept as one leaf. The recurrences are
\begin{align}
 n(U)&=1 &&\text{if $U$ is a leaf},\nonumber\\
 n(U)&=n(V)+n(W)+1 &&\text{if $U$ is an addition node},\label{eq:formula-size-add}\\
 n(U)&=n(V)+n(W)+|V|+2 &&\text{if $U$ is a multiplication node}.\label{eq:formula-size-product}
\end{align}
In the last line, $|V|$ counts the unchanged copy of the original left subtree. The final $2$ counts one multiplication gate and one addition gate used to join the terms in~\eqref{eq:formula-product}.

There is exactly one recursive call for each node occurrence of $\Phi$: at either kind of internal node we call each original child once. In particular, inserting the unchanged copy of the left subtree does not start a further recursion on that copy. Let $\mu$ be the number of original multiplication nodes, and let $V_U$ be the left child of a multiplication node $U$. Summing the recurrences gives
\begin{equation}\label{eq:formula-exact-size}
 n(\rho)=s+\mu+
 \sum_{\substack{U\text{ a multiplication}\\\text{node of }\Phi}}|V_U|.
\end{equation}
To see each term in~\eqref{eq:formula-exact-size}, first count one node for every original node occurrence. This contributes $s$: one constant leaf at an original leaf, one addition gate at an original addition, and one of the two connecting gates at an original multiplication. Each multiplication contributes one further connecting gate, accounting for $\mu$, and its unchanged left-subtree copy, accounting for the sum.

Thus the extra contribution at each multiplication node is $|V_U|+1\le s+1$. There are at most $s$ multiplication nodes. It follows directly that
\[
 n(\rho)\le s+\mu(s+1)=O(s^2).
\]
The final multiplication by $c^{-1}$ adds only a scalar leaf and a multiplication gate. This is a formula-size count with no sharing: every copied occurrence is included.

\emph{Construction time.}
First compute the matrices $C_U$. Ordinary matrix addition and multiplication use $O(sb^3)$ field operations in total. Next start the formula recursion with $z=e_b$. At a multiplication node, computing the new column $C_Wz$ costs $O(b^2)$ field operations. At a variable leaf, evaluating $e_0^{\mathsf T}A_xz$ costs at most the same amount. There are $s$ recursive calls, so this part uses $O(sb^2)$ field operations. Finally, writing the unchanged subtree copies and connecting gates takes $O(s^2)$ operations, by the size count above. These are arithmetic operation bounds; bit costs depend on the representation of the field and its elements.
\end{proof}

\begin{corollary}[Formula factor closure]\label{cor:formula}
If a nonzero polynomial $f$ has a formula of size $s$, each one-sided factor has a formula of size $O(s^2)$. If $f=agb\ne0$, then $g$ has a formula of size $O(s^4)$.
\end{corollary}
\begin{proof}
Apply Lemma~\ref{lem:formula} to the map removing the right cofactor. Reversal preserves formula size, so it gives the same bound for a right factor. For $f=agb$, remove $b$ and then, using reversal, remove $a$. There are two formula transformations. The degree of a polynomial computed by a size-$s$ formula is at most $s$, so the data and arithmetic construction time are polynomial in $s$.
\end{proof}

The scalar computations in this proof are performed while constructing the formula. Their results are supplied as constant leaves. This is valid over every field in our arithmetic model, including $\Q$; Section~\ref{sec:fields} explains why binary lengths are a separate issue.

\subsection{Algebraic branching programs}\label{sec:abp}

An affine layered algebraic branching program (ABP) has a source, a sink, and edges between consecutive layers. Each edge is labelled by an affine linear polynomial. Its value is the sum of the ordered products of labels along source-to-sink paths. Its length is the number of edges on each such path, and its width is the largest layer size.

The ABP case is included for completeness. Polynomial-size ABP factorizations over finite fields are already known from the journal version~\cite{AJ25}, which gives a randomized polynomial-time factorization algorithm with ABP inputs and outputs. The following direct proof shows that the same extraction automaton preserves polynomial ABP size. We state and prove the size bound over every field.

\begin{proposition}[ABP factor closure]\label{prop:abp}
Every factor of a polynomial computed by a polynomial-size noncommutative ABP has a polynomial-size ABP over the same field.
\end{proposition}
\begin{proof}
Let an ABP compute $P$, and put $N=b+2$ for a nonscalar removal map. Replace each vertex by $N$ vertices indexed by the NFA states. An edge labelled $L=a_0+\sum_x a_xx$ is replaced by edges whose labels are the entries of
\[
 L(M)=a_0I_N+\sum_xa_xM_x.
\]
These labels are still affine linear polynomials. For each original path, summing over the new state indices gives the corresponding entry of the ordered product of its label matrices. Summing over original paths therefore gives $P(M)$. Choose state $*$ at the source and state $b$ at the sink, and multiply by $c^{-1}$ on a final constant edge. The resulting ABP computes $T(P)$ by Lemma~\ref{lem:matrix}.

Each original edge gives at most $N^2$ edges, and each layer width grows by at most $N$, apart from the final single sink. Scalar removal uses just a constant edge. Reversing edges and interchanging source and sink reverses every path word, so the construction also removes a left cofactor. Apply it twice for a central factor. The cofactor degrees are bounded by the input ABP length, proving polynomial size.
\end{proof}

\subsection{Formula outputs from finite-field factorization}\label{sec:formula-algorithm}

The size theorem can be combined with an existing algorithm that finds ABP factors. We first explain how to obtain the required cofactor data from an ABP.

\begin{lemma}\label{lem:abp-data}
Given an ABP for a nonzero $h\in\F_q\langle X\rangle$, a word $v$ of degree $\deg h$ with $[v]h\ne0$, and the coefficients of all its suffixes in $h$, can be found in deterministic time polynomial in the encoded ABP size and $\log q$.
\end{lemma}
\begin{proof}
The degree is at most the ABP length. Form its homogeneous components by recording at every vertex the degree used so far, splitting each affine edge into its constant and linear parts. This increases the size only polynomially. Within a fixed degree, eliminate scalar edges by summing the weights of scalar-only paths in the acyclic graph, using dynamic programming. The degree-$j$ component then has a representation
\[
 a^{\mathsf T}L_1(X)\cdots L_j(X)z,
 \qquad L_i(X)=\sum_{x\in X}B_{i,x}x,
\]
where the matrices and endpoint vectors have polynomial dimension and scalar entries. Deterministic ABP identity testing~\cite{RS05} finds the largest nonzero degree $b$ by testing these components.

In the degree-$b$ component, suppose a prefix $w=w_1\cdots w_i$ has been fixed. The polynomial of its possible completions is represented by the remaining layers, with initial row
\[
 a^{\mathsf T}B_{1,w_1}\cdots B_{i,w_i}.
\]
If this polynomial is nonzero and $i<b$, at least one next letter leaves a nonzero completion polynomial. Try the letters and use the same identity test to choose one. After $b$ steps the chosen word $v$ has nonzero coefficient. To compute a suffix coefficient, use the component of that suffix's length and multiply its letter coefficient matrices. This takes polynomial time for all $b+1$ suffixes. If $b=0$, the polynomial is a nonzero scalar and its value is obtained by setting all variables to zero.
\end{proof}

\begin{corollary}[Finite-field factorization with formula outputs]\label{cor:formula-algorithm}
Given a nonzero noncommutative formula of size $s$ over an explicitly represented finite field $\F_q$, a complete factorization can be computed in randomized time polynomial in $s$, the number of variables, and $\log q$. The output is a nonzero scalar and an ordered list of formulas for the nonconstant irreducible factors. Each output formula has size $O(s^4)$.
\end{corollary}
\begin{proof}
Apply the algorithm of~\cite{AJ22} to the input formula. It gives polynomial-size ABPs for a complete factorization $f=\lambda f_1\cdots f_k$. For each $i$, concatenate the earlier and later ABPs to represent
\[
 a_i=\lambda f_1\cdots f_{i-1},\qquad b_i=f_{i+1}\cdots f_k,
 \qquad f=a_if_ib_i.
\]
Empty products are $1$. These cofactor ABPs have polynomial size. Obtain the removal data for $b_i$ and $\rev(a_i)$ using Lemma~\ref{lem:abp-data}; reversing the edges gives an ABP for the reversed cofactor.

Starting from the original formula for $f$, remove $b_i$, reverse, remove $\rev(a_i)$, and reverse again. Lemma~\ref{lem:formula} gives successive formula sizes $O(s^2)$ and $O(s^4)$. Scalar cofactors only require scalar multiplication. We start separately from the original formula for each $i$, so the exponent does not grow along the factor list. Since $k\le\deg f\le s$, there are at most $s$ output factors and at most $O(s^5)$ output formula nodes in total. The cofactor ABPs have polynomial size, their data are recovered in polynomial time, and all scalar computations in Lemma~\ref{lem:formula} have polynomial bit cost over $\F_q$. Thus the whole reconstruction takes polynomial time. A scalar input has an empty list of nonconstant factors and is handled separately.

The reconstruction is deterministic once the ABP factorization is supplied, and preserves its factors exactly. Its success probability is therefore that of the algorithm in~\cite{AJ22}. The conference version~\cite{AJ22} already suffices for this corollary. We use the factor ABPs to obtain cofactor data and transform the original formula; no general conversion of an arbitrary ABP into a small formula is assumed.
\end{proof}

\section{Closure of VP and VNP}\label{sec:classes}

We use the standard nonuniform definitions over a fixed field~\cite{AJR16}. A family has polynomially many variables and polynomial degree. It belongs to $\VPnc$ if it has polynomial-size noncommutative circuits.

\subsection{VP closure}

If $f_n=a_ng_nb_n\ne0$ and $(f_n)\in\VPnc$, Theorem~\ref{thm:size} gives polynomial-size circuits for every specified $g_n$. Degree additivity gives polynomial degree as well. The cofactor data chosen for each $n$ become constants in the output circuit. Their existence suffices for this nonuniform closure statement; finding them is a separate algorithmic problem.

\subsection{VNP closure}

Let $X$ and $Z=\{z_1,\ldots,z_t\}$ be disjoint sets of free noncommuting variables. A circuit for $G(X,Z)\in\F\langle X\cup Z\rangle$, together with the designation of $Z$ as Boolean inputs, represents
\begin{equation}\label{eq:vnp}
 f(X)=\sum_{e\in\{0,1\}^t}G(X,e).
\end{equation}
Here $G(X,e)$ is the polynomial in $X$ obtained by replacing $z_j$ by the field scalar $e_j$. The circuit is for $G(X,Z)$; we do not assume a small circuit for the whole sum $f$. The polynomial $G$ is called a verifier. A family $(f_n)$ is in $\VNPnc$ when the number of variables, the number of Boolean inputs, and the verifier size and degree are polynomially bounded in $n$.

\begin{theorem}[VNP factor closure]\label{thm:vnp}
Over every field, $\VNPnc$ is closed under taking factors. Quantitatively, suppose $f\ne0$ has a representation~\eqref{eq:vnp} whose verifier $G$ has circuit size $s$ and degree at most $d$, where $d\ge1$.
\begin{enumerate}[label=(\roman*),leftmargin=*]
\item If $f=gh\ne0$, then $g=\sum_{e\in\{0,1\}^t}H(X,e)$ for a verifier $H$ of size $O(sd^3)$ and degree at most $d$. The same bounds hold for a right factor.
\item If $f=a q k\ne0$, then $q$ has a verifier of size $O(sd^6)$ and degree at most $d$.
\end{enumerate}
The new verifiers use the same Boolean variables $Z$. A constant verifier is handled by scalar arithmetic.
\end{theorem}
\begin{proof}
We first prove (i). Suppose $f=gh$ and $b=\deg h\ge1$. Choose the cofactor data for $h$ from the factorization of the whole polynomial $f$. Let
\[
 T_h:\F\langle X\rangle\longrightarrow\F\langle X\rangle
\]
be the corresponding linear map of Theorem~\ref{thm:remove}. Its extraction property is $T_h(gh)=g$. The map is defined on every polynomial in $X$, even if that polynomial is not divisible by $h$.

Let $c\ne0$ and $M_x$ be the normalization constant and matrices of Lemma~\ref{lem:matrix} for these same cofactor data. Set $N=b+2$. The matrices are fixed independently of the Boolean assignment, and for every $P\in\F\langle X\rangle$ the lemma gives
\[
 T_h(P)=c^{-1}[P((M_x)_{x\in X})]_{*,b}.
\]
Each entry of $M_x$ has degree at most one in $X$.

Construct a matrix circuit from the circuit for $G(X,Z)$ by substituting
\[
 x\longmapsto M_x,\qquad z_j\longmapsto z_jI_N,\qquad
 \lambda\longmapsto\lambda I_N\quad(\lambda\in\F),
\]
and using matrix addition and ordered matrix multiplication at its gates. Let $\widehat G(X,Z)$ be its output matrix and define
\begin{equation}\label{eq:vnp-verifier}
 H(X,Z)=c^{-1}[\widehat G(X,Z)]_{*,b}
       =c^{-1}\bigl[G((M_x)_x,(z_jI_N)_j)\bigr]_{*,b}.
\end{equation}
This is one verifier circuit in the formal variables $X\cup Z$.

Fix a Boolean assignment $e$. Specialization respects addition and ordered multiplication. Both ways of evaluating the original circuit send $x$ to $M_x$, $z_j$ to $e_jI_N$, and a field constant $\lambda$ to $\lambda I_N$. Induction over the gates therefore gives
\[
 \widehat G(X,e)=G((M_x)_x,(e_jI_N)_j)
              =\bigl(G(X,e)\bigr)((M_x)_x).
\]
Applying Lemma~\ref{lem:matrix} to $G(X,e)$ yields $H(X,e)=T_h(G(X,e))$. Hence
\begin{align}\label{eq:vnp-proof}
 \sum_{e\in\{0,1\}^t}H(X,e)
 &=\sum_{e\in\{0,1\}^t}T_h(G(X,e))\nonumber\\
 &=T_h\!\left(\sum_{e\in\{0,1\}^t}G(X,e)\right)
 =T_h(f)=T_h(gh)=g.
\end{align}
The second equality uses linearity of the fixed map $T_h$; the final equality is Theorem~\ref{thm:remove}. Individual summands need not be divisible by $h$. Before specialization, $z_jI_N$ need not commute with $M_x$, and no such commutation was used.

To implement the verifier by ordinary circuit gates, each matrix addition costs $O(N^2)$ gates and each multiplication costs $O(N^3)$ gates. Sharing previously computed entries gives size $O(sN^3)=O(sb^3)$. Degree additivity and~\eqref{eq:vnp} imply $b\le\deg f\le d$, so this is $O(sd^3)$. Every substituted entry has degree at most one in $X\cup Z$, and therefore $\deg H\le\deg G\le d$. This degree statement concerns the polynomial computed by the verifier and remains valid in the presence of cancellations. A supplied syntactic degree bound is also preserved. If $h=c$ is scalar, take $H=c^{-1}G$.

Interchanging the inputs at every multiplication gate computes word reversal without changing circuit size. Reversal commutes with Boolean specialization, because the assigned values are field scalars. Thus it preserves Boolean-sum representations, and applying (i) to $\rev(f)=\rev(h)\rev(g)$ gives the right-factor claim.

For (ii), first remove $k$ to obtain a verifier for $a q$. Reverse, remove the right cofactor $\rev(a)$ from $\rev(q)\rev(a)$, and reverse again. Both cofactor degrees are at most $d$. Two applications of the preceding construction give size $O(sd^6)$, degree at most $d$, and the same Boolean variables. For a VNP family all these parameters are polynomially bounded, proving closure for every specified factor family.
\end{proof}

The proof uses more than the assertion of VP factor closure: $f$ need not have a small circuit, and its summands need not share a factor. What permits the passage to VNP is one cofactor-dependent linear map together with one small verifier implementing it for all Boolean assignments. The data are chosen nonuniformly for the specified factor. Together with Corollary~\ref{cor:formula} and the VP argument, Theorem~\ref{thm:vnp} proves Theorem~\ref{thm:closure}.

\begin{corollary}\label{cor:multiple}
If a polynomial-degree family $(g_n)$ requires superpolynomial noncommutative circuit size, then every polynomial-degree family $(f_n)$ satisfying $f_n=a_ng_nb_n\ne0$ also requires superpolynomial circuit size. The same statement holds for formulas and for ABPs in their respective size measures.
\end{corollary}
\begin{proof}
Small representations of $(f_n)$ would give small representations of $(g_n)$ by the corresponding closure theorem.
\end{proof}

\section{Succinct ABPs with polynomial-time edge queries}\label{sec:succinct}

For this section and the next, all bit-complexity statements are over an explicitly represented finite field. We use the affine layered ABPs defined in Section~\ref{sec:abp}.

A succinct description gives the layer sizes and an algorithm that, on binary indices $(i,j,k)$, returns the edge label from vertex $i$ of layer $k-1$ to vertex $j$ of layer $k$. A nonedge has label zero. Both the length of this description and the cost of an edge query matter.

\begin{lemma}[Computing one coefficient]\label{lem:coefficient}
For a circuit $C$ and a word $w$ of length $d\ge1$, the coefficient $[w]C$ can be computed in $O(\operatorname{size}(C)d^3)$ field operations. The empty-word coefficient can be computed in $O(\operatorname{size}(C))$ field operations. No degree bound on intermediate gates is needed.
\end{lemma}
\begin{proof}
At each gate, compute the coefficient of every substring $w[a:b]$, including empty substrings. Input gates are immediate. Addition is componentwise. At a product gate use
\[
 [w[a:b]](PQ)=\sum_{t=a}^b[w[a:t]]P\,[w[t:b]]Q.
\]
Process gates in topological order. For $d\ge1$, there are $O(d^2)$ intervals and $O(d)$ cuts for each interval. Longer words in the gate polynomials cannot contribute to these coefficients. The empty-word coefficient is obtained by evaluating the circuit at $X=0$.
\end{proof}

\begin{theorem}[Local representation of a factor]\label{thm:succinct}
Suppose a size-$s$ circuit over $\F_q$ computes a nonzero polynomial $f$ of degree at most $D$ in $m$ variables. Every factor has an affine layered ABP of length $D+1$ and width at most $\sum_{j=0}^D m^j$, whose description has polynomial length in $s,D,m,\log q$ and whose edge labels are computable in deterministic polynomial time in these parameters.
\end{theorem}
\begin{proof}
Let $P$ be the specified factor. For $D\ge1$, Theorem~\ref{thm:size} gives it a circuit of size $\poly(s,D)$; for $D=0$, $P$ is a scalar. For $D\ge1$, make a separate path for each word $w\in X^{\le D}$. Give each path $D$ internal vertices, identify all sources, and identify all sinks. The first edge has label $[w]P$. The remaining $D$ edges spell the letters of $w$ and then have label $1$ until the sink is reached. Thus each path has length $D+1$ and weight $[w]P\,w$, and the ABP computes $P$.

Number the words by length and then lexicographically. An internal vertex is identified by its word number and layer. The word number has $O(\log(D+1)+D\log(m+1))$ bits and can be converted to a word in polynomial time. An edge query checks whether the indices describe consecutive vertices of one path, allowing for the common source and sink. The label is then zero, one, a variable, or the coefficient $[w]P$. The last is computable in polynomial time by Lemma~\ref{lem:coefficient} and finite-field arithmetic. A full affine coefficient list has only $m+1$ entries. For $D=0$, a single constant edge suffices.

The description may include the factor circuit itself. Equivalently, it may include the original circuit and the at most two sets of NFA data from Theorem~\ref{thm:size}, from which that circuit is built in polynomial time. All their field elements have $O(\log q)$ bits.
\end{proof}

This theorem gives a deterministic polynomial-time local evaluator. Finding its description is a separate task. The possibly exponential width is allowed; no minimum-width claim is made. Over $\Q$, the coefficient algorithm still uses polynomially many field operations, but explicitly writing one coefficient can require many bits. The finite-field bit-time conclusion does not follow over $\Q$ from circuit size alone.

\section{Finding a complete factorization over finite fields}\label{sec:search}

Given a circuit over an explicitly represented finite field $\F_q$ for a nonzero polynomial $f$, and a degree bound $D$ in unary, we seek a complete ordered factorization $f=\lambda f_1\cdots f_k$. The output consists of $\lambda\in\F_q^\times$ and circuits over $\F_q$ for the nonconstant irreducible factors $f_i$. The list is empty for a scalar input. We prove a deterministic $\FP^{\SigTwo}$ bound and a zero-error randomized bound with an NP oracle and expected polynomial running time.

Both algorithms use the factor-size theorem to search for bounded circuits $G,H$ giving a nontrivial split of a current factor $F$. Comparing every word coefficient of $F-GH$ gives an existential circuit choice followed by a universal word test, yielding the first bound. For the second, we first sample and certify a polynomial-size list of matrix evaluations that tests every bounded candidate identity exactly. With the list fixed, verifying a proposed split takes polynomial time, so split search requires only an NP oracle.

A degree bound $D=0$ gives a scalar input, handled by evaluation at $X=0$. Assume $D\ge1$. Replace the input by the homogeneous-component construction of Section~\ref{sec:prelim}, of syntactic degree at most $D$, and let $s$ be its size. Fix the original factor bound $B=B(s,D)=KsD^6\ge s$ from Remark~\ref{rem:bound}. We use this same bound throughout the search. All size and degree bounds in oracle queries are encoded in unary and have values polynomial in the original input length and $D$. Relabel the input variables as $X=\{x_1,\ldots,x_m\}$, with $m$ at most the input circuit size. Lemma~\ref{lem:coefficient} computes each specified polynomial-length word coefficient in polynomial time.

\subsection{Deterministic search with a second-level oracle}
\begin{theorem}[Deterministic factorization]\label{thm:search}
Given the circuit and unary degree bound above, a complete ordered irreducible factorization can be computed in $\FP^{\SigTwo}$. The output is a scalar and a list of circuits, with total encoded length polynomial in the input length and $D$.
\end{theorem}
\begin{proof}
First ask whether some nonempty word of length at most $D$ has a nonzero coefficient in $f$. This is an NP question. If not, the input is a nonzero scalar: output $f(0)$ and an empty factor list. Henceforth the current polynomial $F$ is a nonconstant factor of $f$, represented by a circuit of size at most $B$.

A nontrivial split is described by circuits $G,H$, each of size at most $B$ and syntactic degree at most $D$, and nonempty words $u,v$ of length at most $D$, satisfying
\begin{equation}\label{eq:split}
 [u]G\ne0,\qquad [v]H\ne0,\qquad
 \forall w\in X^{\le2D}:\ [w](F-GH)=0.
\end{equation}
The first two conditions certify that both factors are nonconstant. The last certifies equality: $\deg(F-GH)\le2D$, so checking these coefficients checks the whole polynomial.

The circuits and the two nonconstancy words form a polynomial-length existential witness. For any specified $w$, all checks in \eqref{eq:split} take polynomial time. Thus existence of a split is a $\SigTwo$ predicate. Here $\SigTwo$ means one polynomial-length existential quantifier followed by one polynomial-length universal quantifier and a polynomial-time check. Word encodings have polynomial length; malformed encodings can be ignored.

This bounded search misses no genuine split. If $F=GH$ is a nontrivial factorization, both $G$ and $H$ are factors of the original $f$, so the original bound $B$ applies to them. Conversely, every successful witness gives an actual nontrivial split.

If a split exists, recover a witness one bit at a time. At each step, ask whether the proposed prefix can be extended to a successful witness. Fixing some existential bits leaves a $\SigTwo$ predicate. Polynomially many queries recover both circuits. Recurse on them in product order. If no split exists, output $F$ as irreducible.

Each split has two positive-degree factors. Degree additivity therefore bounds the recursion by $D$ leaves and $2D-1$ nodes. Every circuit found obeys the same original bound $B$, so the queries, computation, and total output length remain polynomial. For a nonconstant input we may take $\lambda=1$, absorbing scalar choices into the factors. Fixed encoding and witness order make the selection deterministic.
\end{proof}

\subsection{Zero-error randomized search with an NP oracle}
\begin{theorem}[Zero-error randomized factorization]\label{thm:np-randomized}
Given the circuit and unary degree bound above, a complete ordered irreducible factorization can be computed with an NP oracle by a zero-error randomized algorithm of expected polynomial running time. It terminates with probability one and outputs a scalar and a list of circuits over the input field, of polynomial total encoded length.
\end{theorem}
\begin{proof}
Handle scalar inputs as above, and assume $D\ge1$. Every candidate identity in the preceding search is $F-GH=0$. Fix
\[
 S=4B+10,\qquad L=2D,
\]
and let $\mathcal C_{S,L}$ consist of circuits over $\F_q$, in the input variables $X$, of size at most $S$ and syntactic degree at most $L$. Every candidate difference $F-GH$ belongs to this class.

A finite list $\mathcal H$ of matrix substitutions is called valid if every nonzero polynomial computed by a circuit in $\mathcal C_{S,L}$ has a nonzero value on at least one substitution in the list. For this class, a valid list makes vanishing on all tests equivalent to being the zero polynomial.

\emph{Sampling a list.}
Matrix evaluations give randomized identity tests for noncommutative circuits~\cite{BW05}. For the test used here, choose a finite extension $E/\F_q$ with $|E|\ge4L$ and polynomial-bit representation.\footnote{If $q\ge4L$, use $\F_q$. Otherwise choose the least $e$ with $q^e\ge4L$. Then $q^e<4qL<16L^2$. Enumerating monic degree-$e$ polynomials over $\F_q$ and applying the standard polynomial-time irreducibility test constructs a suitable extension in polynomial time.}
For each variable $x$, independently sample uniform elements $a_{i,x}\in E$, for $0\le i<L$, and substitute
\[
 Z_x=\sum_{i=0}^{L-1}a_{i,x}E_{i,i+1}.
\]
Here $Z_x$ is an $(L+1)\times(L+1)$ matrix, indexed from $0$ to $L$, and $E_{i,i+1}$ has a single $1$ in position $(i,i+1)$ and zeros elsewhere. Thus every letter advances one position.

To analyze a test, first regard the weights $a_{i,x}$ as commuting indeterminates. For a polynomial $P$ of degree at most $L$,
\begin{equation}\label{eq:shift-test}
 [P(Z)]_{0,k}
 =\sum_{x_{j_1}\cdots x_{j_k}\in X^k}
 [x_{j_1}\cdots x_{j_k}]P
 \prod_{i=0}^{k-1}a_{i,x_{j_{i+1}}}
 \qquad(1\le k\le L).
\end{equation}
The diagonal entries equal the constant coefficient. Different words on the right of \eqref{eq:shift-test} give different commuting monomials, because each weight records both its letter and its position. Hence, if $P\ne0$, some matrix entry is a nonzero polynomial in the weights, of degree at most $L$.

After assigning independent uniform field values to the weights, the polynomial zero bound~\cite{Schwartz80} gives
\[
 \Pr[P(Z)=0]\le \frac{L}{|E|}\le\frac14.
\]
Nonzero constants are detected with probability one. A test is performed by evaluating the circuit on these matrices, in polynomial time.

One random test may miss some circuits. We need a list that works simultaneously for all circuits in $\mathcal C_{S,L}$. Fix a padded binary encoding of this class, of length
\[
 \ell=O\bigl(S(\log(S+m+2)+\log q)\bigr).
\]
This accounts for the gate types, wires, input labels, constants, and output gate. There are at most $2^\ell$ encoded circuits. Sample $\ell+2$ independent substitutions. For any fixed nonzero circuit, the probability of missing all its tests is at most $4^{-(\ell+2)}$. A union bound therefore gives
\[
 \Pr[\text{the list is invalid}]
 \le 2^\ell4^{-(\ell+2)}<\frac14.
\]
The list has polynomial length and is valid with probability greater than $3/4$. Random sampling alone does not identify which sampled lists are valid. The next step supplies that guarantee.

\emph{Certifying the list.}
A proposed list $\mathcal H$ is invalid exactly when there exist a circuit $C\in\mathcal C_{S,L}$ and a word $w\in X^{\le L}$ such that
\begin{equation}\label{eq:bad-list}
 [w]C\ne0
 \quad\text{and}\quad
 C(Z)=0\ \text{for every }Z\in\mathcal H.
\end{equation}
The coefficient proves that $C$ computes a nonzero polynomial, while the evaluations show that the list misses it.

This is an NP predicate. The circuit and word are polynomial-length witnesses. Size and syntactic degree are checked directly; Lemma~\ref{lem:coefficient} computes the specified coefficient; and the polynomially many matrix evaluations are polynomial-time computations. Empty words are allowed, so the test also covers a nonzero constant coefficient.

Ask the NP oracle whether \eqref{eq:bad-list} has a witness. If the answer is yes, discard the list and sample a fresh one. If the answer is no, retain it. A retained list is certainly valid. Each independent trial succeeds with probability greater than $3/4$, so the expected number of trials is less than $4/3$, and termination occurs with probability one. Exact uniform sampling of field elements by rejection sampling also has expected polynomial cost.

\emph{Searching for factors.}
Now use the same search as in Theorem~\ref{thm:search}, replacing its universal coefficient test by
\[
 (F-GH)(Z)=0\qquad\text{for every }Z\in\mathcal H.
\]
Because $F-GH\in\mathcal C_{S,L}$ and the list is valid, these checks are equivalent to the polynomial identity $F=GH$.

A split query guesses only $G,H$ within the fixed bounds and words certifying that they are nonconstant. Checking this witness now takes polynomial time, so the split predicate is in NP. Certification of the list was completed beforehand; it is not an additional condition inside this NP verifier.

Recover witnesses by fixing their encoding bits through NP queries, and recurse in product order. The same bound $B$ detects every genuine split, and the same degree argument gives at most $2D-1$ search nodes. All guessed factors have coefficients in the original field $\F_q$; the extension field is used only for evaluating tests.

Fix a padded encoding of the split witnesses, and always try bit $0$ before bit $1$ when recovering a witness. Every certified list accepts exactly the witnesses to a genuine split, so all such lists give the same answers to the prefix queries. Inductively, they yield the same ordered output factorization. Thus the random choices affect the running time, but need not affect which factorization is returned.

The list remains valid for candidates chosen using previous oracle answers: it was certified for the entire bounded class, including every possible candidate difference. Thus all later decisions, including declarations of irreducibility, are exact.

After certification, the computation is deterministic polynomial time with an NP oracle. Only the sampling stage is randomized, and it has zero error and expected polynomial running time. This proves the theorem.
\end{proof}

The randomized upper bound also follows from the general inclusion $\mathsf{MA}\subseteq\mathsf{ZPP}^{\NP}$~\cite[Theorem~5]{GZ11}. Indeed, existence of a bounded split, with any prescribed prefix of its witness encoding, is in $\mathsf{MA}$. The witness supplies the factor circuits and nonconstancy words; the verifier checks the size and syntactic degree bounds and the two nonzero coefficients deterministically, and uses randomized noncommutative identity testing~\cite{BW05} to verify the product. A correct identity always passes, while an incorrect one passes with probability at most $1/3$. The inclusion gives exact expected-polynomial-time decisions for these prefix queries. Witness recovery and recursive splitting use polynomially many queries of polynomial length, so their total expected running time is polynomial. The direct proof above implements the bound through one list of matrix evaluations certified for all candidate identities. In both arguments, the factor-size theorem supplies the polynomial bound on the witnesses.

The second result is an expected-polynomial-time randomized oracle bound. It does not give an unconditional deterministic $\FP^{\NP}$ algorithm. A deterministic polynomial-time identity test would give that stronger conclusion directly, by making the verification of a guessed split polynomial time.

\paragraph{Closing remark.}
The extraction proof uses the automaton to arrange cancellation between accepting runs. The identity-testing constructions in~\cite{AMS10,AJMR19} use automata to isolate a contribution that survives, while the word-removal construction in~\cite{AJR18} selects and erases a prescribed word. Here the reset labels are chosen from the cofactor so that, including its word coefficients, runs on different input words have paired contributions that cancel, leaving precisely the required factor. To our knowledge, this cofactor-dependent pairing for nonhomogeneous factor extraction is new.

The closure proofs construct a specified factor from short cofactor data, while the circuit-factorization algorithms find factors by bounded search and identity testing. Finding the cofactor data efficiently from a general circuit remains an open direction. It would also be useful to reduce the degree dependence of the construction and to obtain bit-complexity bounds over $\Q$ under a suitable normalization of factors.

\section*{Acknowledgments}
We acknowledge the use of OpenAI's ChatGPT (GPT-6 Astra) for discussions during this work. We initially set out to explore the computation of succinct algebraic branching programs for factors. A series of discussions led from this question to the factor-extraction NFA and the subsequent VP, VNP, Formula factor-closure results. ChatGPT also assisted in examining proof details, locating relevant literature, and structuring and editing the write-up. The results and text are the authors’ responsibility.


\begin{thebibliography}{99}
\bibitem{AMS10}
V.~Arvind, P.~Mukhopadhyay, and S.~Srinivasan.
\newblock New results on noncommutative and commutative polynomial identity testing.
\newblock \emph{Computational Complexity}, 19(4):521--558, 2010.
\newblock Preliminary version in CCC 2008.
\newblock \href{https://doi.org/10.1007/s00037-010-0299-8}{doi:10.1007/s00037-010-0299-8}.

\bibitem{AJMR19}
V.~Arvind, P.~S.~Joglekar, P.~Mukhopadhyay, and S.~Raja.
\newblock Randomized polynomial-time identity testing for noncommutative circuits.
\newblock \emph{Theory of Computing}, 15(7):1--36, 2019.
\newblock Preliminary version in STOC 2017.
\newblock \href{https://doi.org/10.4086/toc.2019.v015a007}{doi:10.4086/toc.2019.v015a007}.

\bibitem{AJR18}
V.~Arvind, P.~S.~Joglekar, and G.~Rattan.
\newblock On the complexity of noncommutative polynomial factorization.
\newblock \emph{Information and Computation}, 262:22--39, 2018.
\newblock \href{https://doi.org/10.1016/j.ic.2018.05.009}{doi:10.1016/j.ic.2018.05.009}.

\bibitem{AJR16}
V.~Arvind, P.~S.~Joglekar, and S.~Raja.
\newblock Noncommutative Valiant's classes: Structure and complete problems.
\newblock \emph{ACM Transactions on Computation Theory}, 9(1):3:1--3:29, 2016.
\newblock \href{https://doi.org/10.1145/2956230}{doi:10.1145/2956230}.

\bibitem{AJ22}
V.~Arvind and P.~S.~Joglekar.
\newblock On efficient noncommutative polynomial factorization via Higman linearization.
\newblock In \emph{37th Computational Complexity Conference (CCC 2022)},
volume 234 of \emph{LIPIcs}, pages 12:1--12:22, 2022.
\newblock \href{https://doi.org/10.4230/LIPIcs.CCC.2022.12}{doi:10.4230/LIPIcs.CCC.2022.12}.

\bibitem{AJ25}
V.~Arvind and P.~S.~Joglekar.
\newblock On efficient noncommutative polynomial factorization via Higman linearization.
\newblock \emph{Computational Complexity}, 34, article 10, 2025.
\newblock \href{https://doi.org/10.1007/s00037-025-00272-9}{doi:10.1007/s00037-025-00272-9}.

\bibitem{BKRSS25}
S.~Bhattacharjee, M.~Kumar, S.~S.~Rai, V.~Ramanathan, R.~Saptharishi, and S.~Saraf.
\newblock Closure under factorization from a result of Furstenberg.
\newblock arXiv:2506.23214, 2025.
\newblock \url{https://arxiv.org/abs/2506.23214}.

\bibitem{BW05}
A.~Bogdanov and H.~Wee.
\newblock More on noncommutative polynomial identity testing.
\newblock In \emph{20th IEEE Conference on Computational Complexity}, pages 92--99, 2005.
\newblock \href{https://doi.org/10.1109/CCC.2005.13}{doi:10.1109/CCC.2005.13}.

\bibitem{CKS19}
C.-N.~Chou, M.~Kumar, and N.~Solomon.
\newblock Closure results for polynomial factorization.
\newblock \emph{Theory of Computing}, 15(13):1--34, 2019.
\newblock \href{https://doi.org/10.4086/toc.2019.v015a013}{doi:10.4086/toc.2019.v015a013}.

\bibitem{Kal89}
E.~Kaltofen.
\newblock Factorization of polynomials given by straight-line programs.
\newblock In S.~Micali, editor, \emph{Randomness and Computation}, volume 5
of \emph{Advances in Computing Research}, pages 375--412. JAI Press, 1989.

\bibitem{GZ11}
O.~Goldreich and D.~Zuckerman.
\newblock Another proof that $\mathsf{BPP}\subseteq\mathsf{PH}$ (and more).
\newblock In O.~Goldreich, editor, \emph{Studies in Complexity and Cryptography:
Miscellanea on the Interplay between Randomness and Computation},
volume 6650 of \emph{Lecture Notes in Computer Science}, pages 40--53.
Springer, 2011.
\newblock Preliminary version: ECCC TR97-045, 1997.
\newblock \href{https://doi.org/10.1007/978-3-642-22670-0_6}{doi:10.1007/978-3-642-22670-0\_6}.

\bibitem{RS05}
R.~Raz and A.~Shpilka.
\newblock Deterministic polynomial identity testing in non-commutative models.
\newblock \emph{Computational Complexity}, 14(1):1--19, 2005.
\newblock \href{https://doi.org/10.1007/s00037-005-0188-8}{doi:10.1007/s00037-005-0188-8}.

\bibitem{Schwartz80}
J.~T.~Schwartz.
\newblock Fast probabilistic algorithms for verification of polynomial identities.
\newblock \emph{Journal of the ACM}, 27(4):701--717, 1980.
\newblock \href{https://doi.org/10.1145/322217.322225}{doi:10.1145/322217.322225}.

\bibitem{Sch19}
K.~Schrempf.
\newblock On the factorization of non-commutative polynomials (in free associative algebras).
\newblock \emph{Journal of Symbolic Computation}, 94:126--148, 2019.
\newblock \href{https://doi.org/10.1016/j.jsc.2018.07.004}{doi:10.1016/j.jsc.2018.07.004}.
\end{thebibliography}
\end{document}